\documentclass[a4paper,12pt]{article}

\usepackage[figuresright]{rotating}
\usepackage{amssymb,amsmath}
\usepackage{titlesec}
\usepackage{geometry}
\usepackage[numbers,sort&compress]{natbib}
\RequirePackage[colorlinks,citecolor=blue,urlcolor=blue]{hyperref}
\RequirePackage{graphicx}
\RequirePackage{multirow}

\title{Bayesian Inference of Phenotypic Plasticity of Cancer Cells Based on Dynamic Model for Temporal Cell Proportion Data}
\author{Shuli Chen, Yuman Wang, Da Zhou\thanks{zhouda@xmu.edu.cn} and Jie Hu\thanks{hujiechelsea@xmu.edu.cn} \\
School of Mathematical Science, Xiamen University, Xiamen, Fujian, China}

\begin{document}
\maketitle
\begin{abstract}
Mounting evidence underscores the prevalent hierarchical organization of cancer tissues. At the foundation of this hierarchy reside cancer stem cells, a subset of cells endowed with the pivotal role of engendering the entire cancer tissue through cell differentiation. In recent times, substantial attention has been directed towards the phenomenon of cancer cell plasticity, where the dynamic interconversion between cancer stem cells and non-stem cancer cells has garnered significant interest.
Since the task of detecting cancer cell plasticity from empirical data remains a formidable challenge, we propose a Bayesian statistical framework designed to infer phenotypic plasticity within cancer cells, utilizing temporal data on cancer stem cell proportions. Our approach is grounded in a stochastic model, adept at capturing the dynamic behaviors of cells. Leveraging Bayesian analysis, we explore the moment equation governing cancer stem cell proportions, derived from the Kolmogorov forward equation of our stochastic model. With improved Euler method for ordinary differential equations, a new statistical method for parameter estimation in nonlinear ordinary differential equations models is developed, which also provides novel ideas for the study of compositional data. Extensive simulations robustly validate the efficacy of our proposed method. To further corroborate our findings, we apply our approach to analyze published data from SW620 colon cancer cell lines. Our results harmonize with \emph{in situ} experiments, thereby reinforcing the utility of our method in discerning and quantifying phenotypic plasticity within cancer cells.

\end{abstract}


\newpage
\section{Introduction}
\label{s:intro}

In multicellular organisms, numerous tissues undergoing frequent cell turnover exhibit a hierarchical architecture. At the core of this cellular hierarchy reside tissue-specific stem cells (SCs), distinguished by their ability to self-renew and differentiate into more mature cell types. It is frequently postulated that analogous hierarchical architectures are present in certain cancerous tissues \citep{reya2001StemCellsCancer}. Analogous to SCs in normal tissues, cancer stem cells (CSCs) possess the capability to undergo unlimited proliferation and differentiate into other non-stem cancer cells (NSCCs) \citep{meacham2013TumourHeterogeneityCancer,batlle2017CancerStemCells}. The CSC hypothesis suggests a unidirectional cascade from CSCs to NSCCs, implying that CSCs can give rise to NSCCs through cellular differentiation, but the reverse is not true \citep{bonnet1997HumanAcuteMyeloid}. Nevertheless, an expanding body of research has begun to challenge this hierarchical model of cancer cells \citep{marjanovic2013CellPlasticityHeterogeneity}. Instances have been documented wherein NSCCs can reclaim CSC-like traits in various cancers, including breast cancer \citep{meyer2009DynamicRegulationCD24,sousa2019HeterogeneityPlasticityBreast}, melanoma \citep{quintana2010PhenotypicHeterogeneityTumorigenic}, colon cancer \citep{yang2012DynamicEquilibriumCancer}, glioblastoma multiforme \citep{chen2012RestrictedCellPopulationa}, and esophageal cancer \citep{li2019ExosomalFMR1AS1Facilitates}. This indicates that the phenotypic transition of cancer cells may indeed be bidirectional rather than unidirectional. Such a bidirectional transition between cells is often referred to as \emph{phenotypic plasticity}. Importantly, the hypotheses of cellular hierarchy and phenotypic plasticity carry distinct implications for cancer therapy \citep{tang2020CancerStemCellsc}. As a result, the effective identification of phenotypic plasticity based on experimental or clinical data emerges as a crucial concern.

In the past decade, computational biology has witnessed a surge in developing mathematical models to quantify cancer cell phenotypic plasticity. Noteworthy examples include \cite{gupta2011StochasticStateTransitions}, who utilized a Markov chain model to predict equilibrium proportions in cancer cell populations by fitting data from breast cancer cell lines. \cite{chen2016OvershootPhenotypicEquilibrium} employed ordinary differential equations (ODEs) to explore the influence of phenotypic plasticity on transient dynamics of cancer stem cells.  \cite{dhawan2016MathematicalModellingPhenotypic} integrated hypoxia into a model, capturing cell plasticity and heterogeneity in transformed mammary epithelial cells. \cite{zhou2018BayesianStatisticalAnalysis} introduced Bayesian analysis for inferring phenotypic plasticity, while \cite{zhou2019InvasionDedifferentiatingCancera} investigated dedifferentiation within the cellular hierarchy. Although models mentioned above have provided valuable insight in understanding the phenotypic plasticity of cancer cells, there are limitations in these works. For example,
\cite{gupta2011StochasticStateTransitions} only adopted the mean value of several trajectories, while the variance information of trajectories
was ignored. \cite{zhou2018BayesianStatisticalAnalysis} proposed a discrete-time Markov chain model, which was obviously against the fact that the division times of cells are stochastic and asynchronous \citep{alt1987StochasticModelCell}. While the others lacked statistical modeling framework of empirical data for characterizing phenotypic plasticity, which were not favorable for statistical inference. Hence, it is necessary to propose a new statistical model composed of different cell division patterns, which can be used for statistical inference of phenotypic plasticity.

In this context, it is worth noting that branching processes have gained extensive utility in characterizing cell proportion dynamics, presenting a rational and efficacious mathematical framework for investigating diverse cell division patterns. Notably, the exploration of statistical inference for cellular branching processes has a longstanding history \citep{tusnady1992StatisticalInferenceBranchingProcessesGuttorp}. However, the proliferation of novel data types has kindled the development of innovative statistical methodologies. While complete genealogy trees originating from underlying branching processes are seldom observable, barring certain exceptions such as the \emph{C. elegans} embryogenesis \citep{hu2015BayesianDetectionEmbryonic}, the observation of the entire genealogy is notably challenging. More frequently, only absolute counts or relative frequencies of distinct cell types are attainable at specific instances. Relative frequencies, being comparatively simpler to record than absolute counts, also offer a valuable avenue for informative extraction. In this context,  the data used in this work is the CSCs proportion in cell populations of SW620 colon cancer cell line \citep{yang2012DynamicEquilibriumCancer}. These proportions, captured at subsequent time points, correspond to what is termed compositional data within the realm of statistics. In this study, CSC plasticity can be considered a form of inter-component dynamics within a cancer cell population, which refer to the interactions, transitions, and changes between different components or subpopulations within a system. However, in the field of statistical analysis for compositional data, the predominant focus of longitudinal compositional data like microbiomics data analyses has revolved around establishing connections between shifts in the compositions and various outcomes \citep{larosa2014PatternedProgressionBacterial,luna2020JointModelingApproach}, with comparatively limited attention directed towards comprehending inter-component dynamics. In the realm of studies investigating dynamic shifts within microbiome data, the generalized Lotka-Volterra model has emerged as a prevalent choice \citep{chung2017IdentificationMicrobiotaDynamics,gibson2018RobustScalableModels}. This model, rooted in predator-prey dynamics, employs kinetic modeling of populations to investigate fluctuations across distinct species. \cite{lugo-martinez2019DynamicInteractionNetwork} also embarked on an endeavor to unearth the processes governing dynamic changes between microbiomes through the application of a dynamic Bayesian network. However, our research does not involve the predator-prey relationship. Moreover, the approach proposed by \cite{lugo-martinez2019DynamicInteractionNetwork} mandates a substantial volume of data for model training, which contrasts with the limited dataset in this study. Thus, given the limitations of existing methods, it is imperative to explore the development of novel statistical techniques for analyzing temporal compositional data in conjunction with the dynamic behavior of cells.

Furthermore, considering the compositional nature of the acquired data, an endeavor can be undertaken to establish a relationship between this compositional data and the original count data. As it is a natural thought to derive the ODEs governing the dynamic change of number of cells by the Kolmogorov forward equation of the branching model, the ODEs on the change of the mean and variance of CSCs proportion can also be formulated. Consequently, our focus shifts towards parameter estimation and model selection within nonlinear ODE models. There are a number of methodologies in this realm, including the nonlinear least squares method (NLS) \citep{xue2010SieveEstimationConstant}, the two-stage smoothing method \citep{brunel2008ParameterEstimationODE}, principal differential analysis (PDA) \citep{ramsay2007ParameterEstimationDifferential}, Bayesian approaches \citep{huang2006HierarchicalBayesianMethods} and the network reconstruction via dynamic systems (NeRDS) method \citep{henderson2014NetworkReconstructionUsing} along with its enhancements \citep{chen2017NetworkReconstructionHighDimensional}.
In the NLS method, the four-stage Runge-Kutta algorithm \citep{runge1895UeberNumerischeAuflosung} is harnessed for numerical solution of the ODEs, facilitating parameter estimation albeit yielding only point estimates. The two-stage smoothing method and PDA necessitate abundant observations or intricate optimization for achieving desired precision levels. Within the current Bayesian paradigm, a likelihood centered on the numerical ODE solution is employed. However, this numerical solution lacks a closed-form, resulting in computationally intensive calculations for each iteration of the sampling process. Meanwhile, the NeRDS method and its extensions assume additivity in the right-hand side of ODEs, an assumption not applicable here.
In response, a new Bayesian framework for this problem is proposed in combination with improved Euler approach, in which the parameter estimation and model selection of interest can be well addressed.

In Section~\ref{s:Method} the real data, model and Bayesian framework are proposed. Main results including simulations and real data analysis are shown in Section~\ref{sec:Results}. Discussions are presented in Section~\ref{sec:discussions}.
\section{Method}
\label{s:Method}
\subsection{Experiment and Data}
\begin{figure}[ht]
 \centerline{\includegraphics[width=15cm]{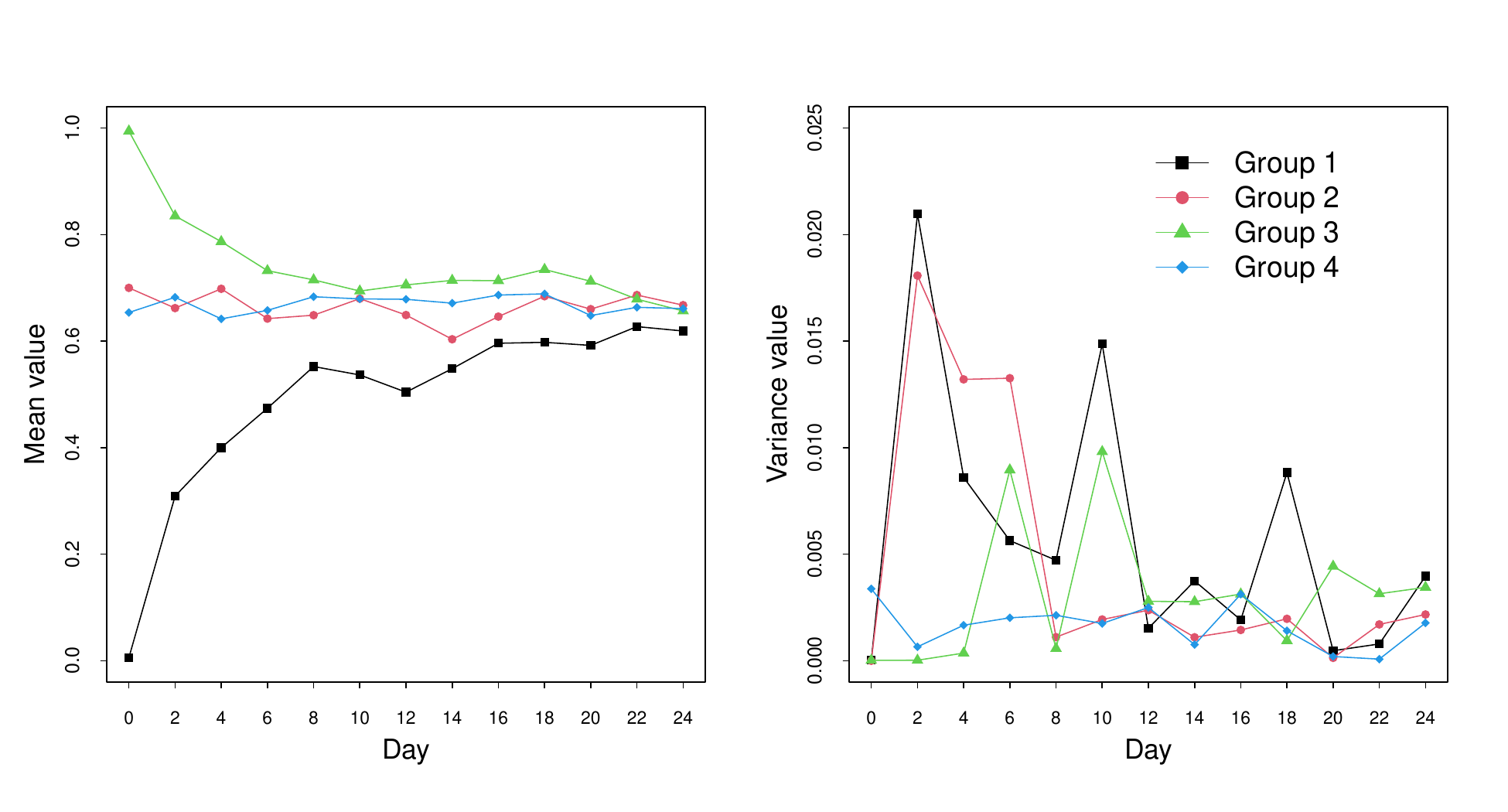}}
\caption{SW620 colon cancer cell line dataset published on \protect\cite{yang2012DynamicEquilibriumCancer}. The left panel and right panel show the temporal mean and variance of CSCs proportion in four different groups, which differ in the initial states of relative frequencies.}
\label{fig:first}
\end{figure}

The dataset applied in this work is the Fluorescence-activated Cell Sorting(FACS) data of SW620 colon cancer cell line, in which two types of cancer cells, i.e. CSCs and NSCCs, were identified. The CSCs and NSCCs were enriched via CD133 cell-surface antigen marker and were isolated using FACS \citep{obrien2007HumanColonCancer}. This dataset consists of four groups, which differ in the initial states of relative frequencies, as shown in Figure~\ref{fig:first}. In each group, both sample mean $m_t$ and sample variance $v_t$ of five trajectories with the same initial state were recorded every second day. The total observation period is 25 days, that is, the data obtained looks like
\begin{eqnarray*}
	\left\{(m_0,v_0),(m_2,v_2),(m_4,v_4),\cdots,(m_{t},v_{t}), \cdots,(m_{24},v_{24})\right\},
\end{eqnarray*}
where $t$ represents observational time point.

\subsection{Model Establishment}\label{subsec:Establishment}
\begin{figure}[t]
 \centerline{\includegraphics[width=12cm]{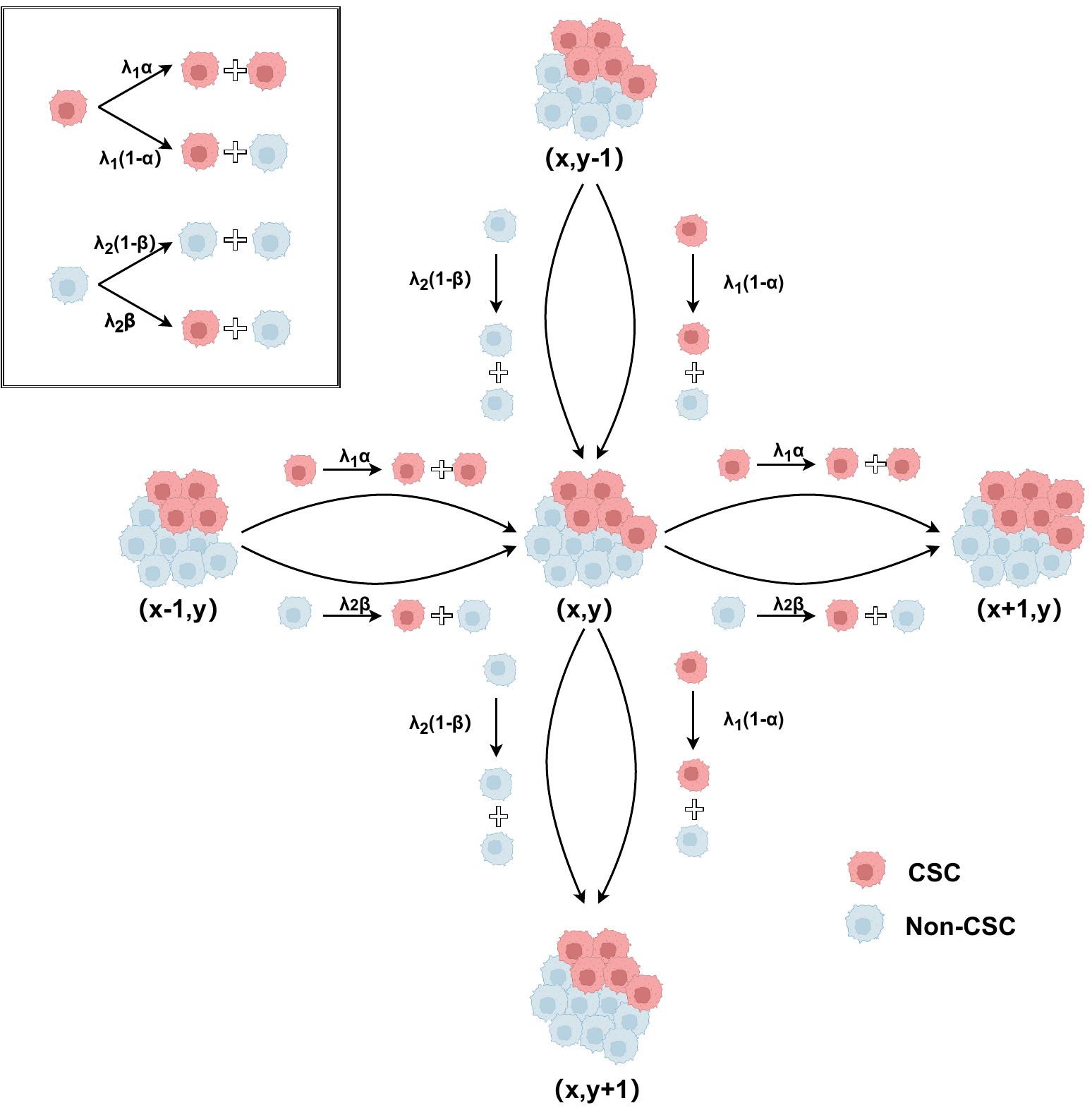}}
\caption{Illustration of the cell-state transitions in our branching model. The upper left corner of the figure illustrates four distinct types of cell divisions. The possible states and processes at adjacent moments are indicated when the number of CSCs and NSCCs at the current moment is x and y respectively.}
\label{fig:third}
\end{figure}

Here a continuous-time branching processes is put forward  to capture the two-type cancer cell population dynamics.
Each CSC is assumed to divide at rate $\lambda_1$, i.e, the waiting time for each cell division event follows exponential distribution with parameter $\lambda_1$. It can either undergo symmetric division giving birth to two identical CSC daughter cells with probability $\alpha$, or undergo asymmetric division giving birth to one CSC daughter cell and one NSCC daughter cell with probability $1-\alpha$. That is, each CSC performs symmetric division at rate $\lambda_1\alpha$ and performs asymmetric division at rate $\lambda_1(1-\alpha)$ (see Figure~\ref{fig:third}).
For each NSCC, it is also assumed that either it divides symmetrically into two NSCC daughter cells with probability $1-\beta$, or it divides asymmetrically into one NSCC daughter cell and one CSC daughter cell with probability $\beta$ (de-differentiation). While assuming each NSCC divides at rate $\lambda_2$, it performs symmetric division at rate $\lambda_2(1-\beta)$ and asymmetric division at rate $\lambda_2\beta$.  When $\beta = 0$, de-differentiation cannot happen, and then the model will reduce to conventional CSCs model. Hence the model selection with respect to $\beta$ provides an efficient way to evaluate whether the phenotypic plasticity is more in accordance with actual situation.

Let $X_t$ be the number of CSC at time $t$, and $Y_t$ be the number of NSCC at time $t$ in one cell line trajectory. Define
\begin{eqnarray*}
	P(x,y,t)=\text{Prob} \{X_t=x, Y_t=y\},
\end{eqnarray*}
then the Kolmogorov forward equation of the branching process is obtained as follow \citep{vankampen2007ChapterMASTEREQUATION}:
\begin{eqnarray}\label{eq:master equation}
	\frac{dP(x,y,t)}{dt}
	&=&\!\!P(x-1,y,t)\lambda_1\alpha(x-1)+P(x-1,y,t)\lambda_2\beta y\nonumber\\
	&\,&\!\!\!\!+P(x,y-1,t)\lambda_1(1-\alpha)x+P(x,y-1,t)\lambda_2(1-\beta)(y-1)\nonumber\\
	&\,&\!\!\!\!-P(x,y,t)(x\lambda_1+y\lambda_2).
\end{eqnarray}
Figure~\ref{fig:third} illustrates the cell-state transition processes of the above Kolmogorov forward equation. To be specific, if it is assumed that the number of CSCs and NSCCs at current moment $t$ is $x$ and $y$ respectively, the last state may be $x-1$ and $y$, or $x$ and $y-1$. And if there were $x-1$ CSCs and $y$ NSCCs at the last state, two types of cell division may happen: one is a symmetric division of CSC giving birth to two identical CSC daughter cells
with rate $\lambda_1 \alpha$, the other is an asymmetrical division of NSCC giving birth to one NSCC daughter cell and one CSC daughter cell with rate $\lambda_2 \beta$, i.e., $P(x-1,y,t)\lambda_1\alpha(x-1)+P(x-1,y,t)\lambda_2\beta y$. Similarly, we can obtain other terms in Equation~(\ref{eq:master equation}). Then the ODE governing the first moment of $X_t$ can be obtained as follows
\begin{align}
	\frac{d\text{E}X_t}{dt}
	&=\sum_{x,y} \frac{x dP(x,y,t)}{dt}\nonumber\\
	&=\lambda_1\alpha \sum_{x,y} P(x-1,y,t)x(x-1)+\lambda_2\beta\sum_{x,y} P(x-1,y,t)x y \nonumber\\ 
	&\quad+\lambda_1(1-\alpha)\sum_{x,y} P(x,y-1,t)x^2 -\sum_{x,y} P(x,y,t)(x\lambda_1+y\lambda_2)x\nonumber\\
	&\quad+\lambda_2(1-\beta) \sum_{x,y} P(x,y-1,t)x(y-1) \nonumber\\
	&=\lambda_1\alpha \sum_{x,y} P(x,y,t)x(x+1)+\lambda_2\beta\sum_{x,y} P(x,y,t)(x+1) y \nonumber\\
	&\quad+\lambda_1(1-\alpha)\sum_{x,y} P(x,y,t)x^2+\lambda_2(1-\beta) \sum_{x,y} P(x,y,t)x y \nonumber\\
	&\quad-\sum_{x,y} P(x,y,t)(x^2\lambda_1+x y\lambda_2)  \quad \mbox{(variable substitution)} \nonumber\\
	&=\sum_{x,y} P(x,y,t)(\lambda_1\alpha x+\lambda_2\beta y)\nonumber\\
	&=\lambda_1\alpha \text{E}X_t+\lambda_2\beta \text{E}Y_t.\label{eq:dEx_t}
\end{align}
And similarly, we can obtain the second moment of $X_t$ (see Web Appendix A for more details)
\begin{eqnarray}\label{eq:second moment}
	\frac{d\text{E}X^2_t}{dt}&=&\sum_{x,y} \frac{x^2 dP(x,y,t)}{dt} \nonumber\\
  &=&\lambda_1\alpha (\text{E}X_t+2\text{E}X^2_t)+\lambda_2\beta (\text{E}Y_t+2\text{E}X_tY_t).
\end{eqnarray}
Then let
\begin{eqnarray*}
	R^X(t)=\frac{X_t}{N_t}
\end{eqnarray*}
be the proportion of CSCs at time $t$,
\begin{eqnarray*}
R^Y(t)=1-R^X(t)
\end{eqnarray*}
be the proportion of NSCCs at time $t$, and
\begin{eqnarray*}
	N_t=X_t+Y_t
\end{eqnarray*}
be the total number of cells in the whole population. $\mu_t$ and $\sigma_t^2$ are introduced to respectively denote the expectation and variance of $R^X(t)$, which can be expressed as
\begin{eqnarray*}
	\mu_t=\text{E}(R^X(t)), \sigma_t^2= \text{Var}(R^X(t)).
\end{eqnarray*}
Since the volatility of $N_t$ is far lower than either $X_t$ or $Y_t$, it is reasonable to assume $N_t$ to be a deterministic variable \citep{gillespie2007StochasticSimulationChemical,yakovlev2009RelativeFrequenciesMultitype}, which is also demonstrated in Section~\ref{subsec:Modification}. Then the ODE of $\mu_t$ can be derived as follows
\begin{eqnarray}
	&\quad&\mu_t = \text{E}(\frac{X_t}{N_t}) \approx \frac{\text{E}X_t}{N_t}\nonumber\label{expectation}\\
	&\Longrightarrow&  \frac{d\text{E}X_t}{dt}=\frac{d(\mu_t N_t)}{dt}=\mu_t \frac{dN_t}{dt}+N_t \frac{d\mu_t}{dt}\nonumber\\
	&\Longrightarrow& \lambda_1\alpha \text{E}X_t+\lambda_2\beta \text{E}Y_t=\mu_t \frac{dN_t}{dt}+N_t \frac{d\mu_t}{dt}\quad \mbox{(substituted by Equation~(\ref{eq:dEx_t}) )}\nonumber\\
	&\Longrightarrow& \lambda_1\alpha\mu_t+\lambda_2\beta(1-\mu_t)=\frac{1}{N_t}\frac{dN_t}{dt}\mu_t+\frac{d\mu_t}{dt}  \quad\mbox{(divided by $N_t$ in both sides)}\nonumber\\
	&\Longrightarrow& \frac{d\mu_t}{dt}=(\lambda_1\alpha-\lambda_2\beta-\frac{1}{N_t}\frac{dN_t}{dt})\mu_t+\lambda_2\beta.\label{eq:dmut}
\end{eqnarray}
Also, approximately we have
\begin{equation}\label{eq:var_appro}
	\sigma^2_t= \text{Var}(\frac{X_t}{N_t}) = \text{E}(\frac{X^2_t}{N^2_t})-\text{E}^2(\frac{X_t}{N_t}) \approx \frac{\text{E}X^2_t}{N^2_t}-\mu^2_t.
\end{equation}
After that, combined with Equation~(\ref{eq:second moment}), the ODE of $\sigma^2_t$ can be obtained as follows (see Web Appendix A for more details)
\begin{eqnarray}\label{eq:dsigma2t} \frac{d\sigma^2_t}{dt}=2(\lambda_1\alpha-\lambda_2\beta-\frac{1}{N_t}\frac{dN_t}{dt})\sigma^2_t+\frac{\lambda_1\alpha-\lambda_2\beta}{N_t}\mu_t+\frac{\lambda_2\beta}{N_t}.
\end{eqnarray}
Meanwhile, let
$\gamma_t$ and $s^{2}_t$ be the expectation and variance of the frequency of NSCCs at time $t$, i.e.
\begin{eqnarray*}
	\gamma_t=\text{E}(R^Y(t)),s^{2}_t=\text{Var}(R^Y(t)).
\end{eqnarray*}
With the same derivation as CSC, the ODEs for NSCCs can be obtained as follows:
\begin{eqnarray}
\frac{d\gamma_t}{dt}\!&=&\![\lambda_2(1\!-\!\beta)\!-\!\lambda_1(1\!-\!\alpha)\!-\!\frac{1}{N_t}\frac{dN_t}{dt}]\gamma_t\!+\!\lambda_1(1\!-\!\alpha) \nonumber\\
\frac{ds^{2}_t}{dt}\!&=&\!2[\lambda_2(1\!-\!\beta)\!-\!\lambda_1(1\!-\!\alpha)\!-\!\frac{1}{N_t}\frac{dN_t}{dt}]s^{2}_t\!+\!\frac{\lambda_2(1-\beta)-\lambda_1(1-\alpha)}{N_t}\gamma_t\!+\!\frac{\lambda_1(1-\alpha)}{N_t}.\label{eq:ds2t}
\end{eqnarray}
\subsection{Model Modification}\label{subsec:Modification}
\begin{figure}[ht]
 \centerline{\includegraphics[width=14cm]{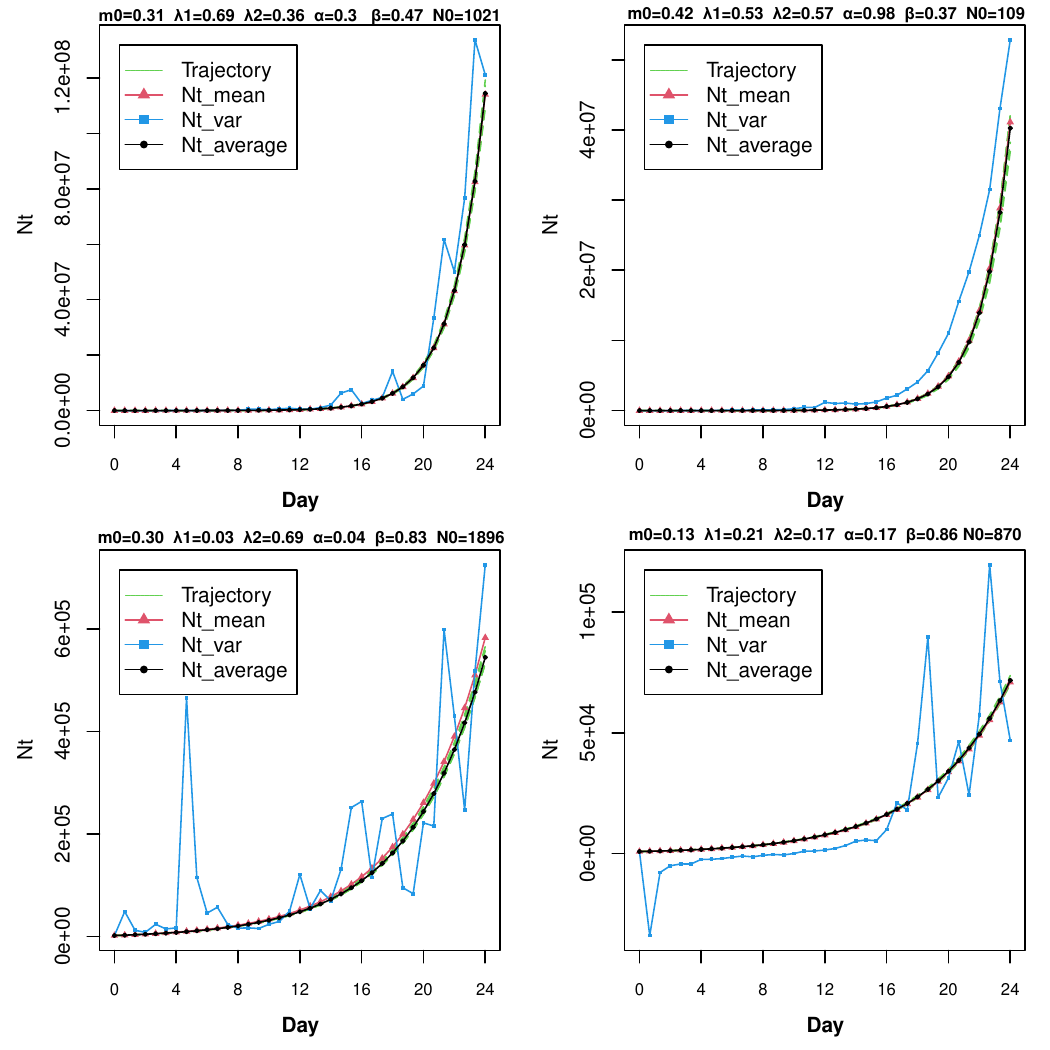}}
\caption{Comparison of simulation results based on Equations~(\ref{eq:Ntmean}) and (\ref{eq:Ntvar}). Green dashed lines represent the total cell numbers generated by Gillespie's algorithm. The black lines named Nt\underline{~}average represent their averaged values. The red lines represent the numbers generated by Equation~(\ref{eq:Ntmean}) and the blue lines represent the counterpart generated by Equation~(\ref{eq:Ntvar}), which are named Nt\underline{~}mean and Nt\underline{~}var respectively.}
\label{fig:forth}
\end{figure}

Note that $\gamma_t=1-\mu_t$, two restraints of $N_t$ should be satisfied, that is
\begin{eqnarray}
&\!\!\!\!\!\!\!\!\!&\frac{d\mu_t}{dt}\!\!=\!\! -\frac{d\gamma_t}{dt} \Rightarrow N_t\!\!=\!\!N_0e^{(\lambda_1-\lambda_2)\int^t _0 \mu_s ds+\lambda_2 t}\label{eq:Ntmean}\\
&\!\!\!\!\!\!\!\!\!&\frac{d\sigma^2_t}{dt}\!\!=\!\!\frac{ds^{2}_t}{dt} \Rightarrow N_t\!\!=\!\frac{(2\lambda_1\alpha\!-\!2\lambda_2\beta\!+\!\lambda_2\!\!-\!\!\lambda_1)\mu_t\!\!+\!\!\lambda_2(2\beta\!\!-\!\!1)}{2(\lambda_2\!-\!\lambda_1)\sigma^2_t}.\label{eq:Ntvar}
\end{eqnarray}
Equations~(\ref{eq:Ntmean}) and (\ref{eq:Ntvar}) provide two different formulaes of $N_t$.
Figure~\ref{fig:forth} shows the comparison of them by simulating $N_t$ in different methods. The initial value $N_0$ and the parameters $\alpha$, $\beta$, $\lambda_1$, and $\lambda_2$ are randomly generated and set differently for each of these four panels. Since $\mu_s$ in Equation~(\ref{eq:Ntmean}) is unknown, the integral can be approximated with discrete observations by a discrete summation, i.e.
\begin{equation}\label{eq:discrete Nt}
	N_{t_k}\approx N_0e^{(\lambda_1-\lambda_2)\sum\limits_{q=0}^{k-1} \frac{m_{t_q}+m_{t_{q+1}}}{2}\Delta t+\lambda_2 t_k},
\end{equation}
here $t_0=0, t_{q+1}= t_q+\Delta t, q=0,1,\cdots,k-1 $, $m_t$ denotes the observation of sample mean of several $R^X(t)$.
From Figure~\ref{fig:forth}, we can see that difference between $N_t$ in various trajectories is small, which indicates that it is reasonable to set $N_t$ as deterministic variable. Since the common approximation methods in Section~\ref{subsec:Establishment} is based on the first moment, which have larger uncertainty for the second moment, the simulation result based on Equation~(\ref{eq:Ntmean}) is closer to the growth of cells. Thus, for more precision, we have
\[	\frac{d\mu_t}{dt}=(\lambda_2-\lambda_1)\mu^2_t+[\lambda_1\alpha-\lambda_2(1+\beta)]\mu_t+\lambda_2\beta  \]
by substituting $N_t$ with Equation~(\ref{eq:Ntmean}) in Equation~(\ref{eq:dmut}).
Due to the deviation induced by our approximation method, if the $N_t$ with form of exponential growth is chosen, the solution of Equation~(\ref{eq:dsigma2t}) does not equal to that of Equation~(\ref{eq:ds2t}). In order to improve the
quality of variance description, we take the average of the right sides of Equations~(\ref{eq:dsigma2t}) and~(\ref{eq:ds2t}) and update the ODE describing the change of $\sigma^2_t$ as follows
\begin{eqnarray*}
	\frac{d\sigma^2_t}{dt}\!\!=\!\![2(\lambda_1\alpha\!\!-\!\!\lambda_2\beta)\!\!-\!\!(\lambda_1\!\!+\!\!\lambda_2)]\sigma^2_t\!\!+\!\!2(\lambda_2\!\!-\!\!\lambda_1)\mu_t\sigma^2_t 
\!\!+\!\!\frac{\lambda_1\!\!-\!\!\lambda_2}{2N_t}\mu_t\!\!+\!\!\frac{\lambda_2}{2N_t}.
\end{eqnarray*}
In summary, a system of ODEs governing the expectation and variance of CSCs proportion is acquired as follows
\begin{equation}
	\left\{
	\begin{array}{ll}
		\frac{d\mu_t}{dt}&\!\!\!=(\lambda_2\!-\!\lambda_1)\mu^2_t\!+\![\lambda_1\alpha\!-\!\lambda_2(1\!+\!\beta)]\mu_t\!+\!\lambda_2\beta \\
		\frac{d\sigma^2_t}{dt}&\!\!\!=[2(\lambda_1\alpha\!\!-\!\!\lambda_2\beta)\!\!-\!\!(\lambda_1\!\!+\!\!\lambda_2)]\sigma^2_t\!\!+\!\!2(\lambda_2-\lambda_1)\mu_t\sigma^2_t
        \!\!+\!\!\frac{\lambda_1\!-\!\lambda_2}{2N_t}\mu_t\!\!+\!\!\frac{\lambda_2}{2N_t},
\end{array}
	\right.
	\label{eq:keyODEs}
\end{equation}
with Equation~(\ref{eq:Ntmean}) holds.
Based on Equation~(\ref{eq:keyODEs}), a Bayesian statistical method for inferring the phenotypic plasticity of cancer cells is designed in the next section.
\subsection{Bayesian Framework}\label{sub:Bayesian framework}
In order to match discrete data points, improved Euler method \citep{gautschi2012NumericalAnalysis} can be used to discretize the ODEs. Let $\{t_k\}$ be a series of time points which satisfy that $t_{k+1}=t_k+\Delta t$. Here $\Delta t$ is a fixed step length, so Equation~(\ref{eq:keyODEs}) can be discretized by two steps. The first is the predictor step as Equation~(\ref{Eq:first step}), starting from the current value and calculating an initial estimation.
\begin{equation}\label{Eq:first step}
\left\{
	\begin{array}{ll}
	\hat{\mu}_{t_{k+1}}\!\!\!\!&\!=\mu_{t_k}+(\lambda_2-\lambda_1)\Delta t \mu^2_{t_k}+[\lambda_1\alpha-\lambda_2(1+\beta)]\Delta t\mu_{t_k}+\lambda_2\beta \Delta t \\
		\hat{\sigma}^2_{t_{k\!+\!1}}\!\!\!\!&\!=\sigma^2_{t_k}\!\!+[2(\lambda_1\alpha\!\!-\!\!\lambda_2\beta)\!\!-\!\!(\lambda_1\!\!+\!\!\lambda_2)]\Delta t\sigma^2_{t_k}\!\!+\!\!2(\lambda_2\!\!-\!\!\lambda_1)\Delta t \mu_{t_k}\sigma^2_{t_k}
		+ \frac{(\lambda_1\!-\!\lambda_2) \Delta t}{2N_{t_k}}\mu_{t_k}\!+\!\frac{\lambda_2 \Delta t}{2N_{t_k}}.\\
	\end{array}
	\right.
\end{equation}
The next step is to replace the value of the next moment at the right end of the implicit Euler format with the initial estimate described above, obtaining the corrected value as follows
\begin{equation}\label{Eq:second step}
	\left\{
	\begin{array}{ll} \bar{\mu}_{t_{k+1}}\!\!\!\!&\!\!\!\!=\mu_{t_k}\!\!+\!\!(\lambda_2\!\!-\!\!\lambda_1)\Delta t \hat{\mu}^2_{t_{k+1}}\!\!\!+\!\![\lambda_1\alpha\!\!-\!\!\lambda_2(1\!\!+\!\!\beta)]\Delta t\hat{\mu}_{t_{k+1}}\!\!\!+\!\!\lambda_2\beta \Delta t \\ \bar{\sigma}^2_{t_{k\!+\!1}}\!\!\!\!&\!\!\!\!=\sigma^2_{t_k}+[2(\lambda_1\alpha\!-\!\lambda_2\beta)\!-\!(\lambda_1\!+\!\lambda_2)]\Delta t\hat{\sigma}^2_{t_{k+1}}\!\!\!
+\!2(\lambda_2\!-\!\lambda_1)\Delta t \hat{\mu}_{t_{k+1}}\hat{\sigma}^2_{t_{k+1}}
		\!+\!\frac{(\lambda_1\!-\!\lambda_2) \Delta t}{2N_{t_{k+1}}}\hat{\mu}_{t_{k+1}}\!+\!\frac{\lambda_2 \Delta t}{2N_{t_{k+1}}}.\nonumber\\
	\end{array}
	\right.\\
\end{equation}
Thus, the difference equation can be considered as an average of the two values above and is expressed as Equation~(\ref{eq:different ode}).
\begin{equation}
\left\{
	\begin{array}{ll} \mu_{t_{k+1}}\!\!\!\!&=\frac{1}{2}(\hat{\mu}_{t_{k+1}}+\bar{\mu}_{t_{k+1}}) \triangleq A_{t_k}(\alpha,\beta,\lambda_1,\lambda_2,\mu_{t_k})\\
\sigma^2_{t_{k\!+\!1}}\!\!\!\!&=\frac{1}{2}(\hat{\sigma}^2_{t_{k\!+\!1}}+\bar{\sigma}^2_{t_{k\!+\!1}})\triangleq \Sigma_{t_k,t_{k+1}}(\alpha,\beta,\lambda_1,\lambda_2,\mu_{t_k},\sigma^2_{t_k},N_{t_{k}},N_{t_{k+1}}).\label{eq:different ode} \\
	\end{array}
	\right.
\end{equation}

The aforementioned derivations are rooted in the context of a single trajectory. Suppose that the sample mean and variance of several trajectories were measured at time $\{t_0,t_1,\cdots,t_K\}$, i.e.
\begin{eqnarray*}
	\left\{(m_{t_0},v_{t_0}),(m_{t_1},v_{t_1}),\cdots,(m_{t_K},v_{t_K})\right\},
\end{eqnarray*}
where $m_{t_k}$ and $v_{t_k}$ denote the observations of sample mean $M_{t_k}$ and sample variance $V_{t_k}$ of $R^X(t_k)$ respectively. By using the Markov property of the branching process, i.e., $P(\mu_{t_{k+1}}=j|\mu_{t_k}=i,\mu_{t_l}=i_l,l=0,1,2\cdots k-1)=P(\mu_{t_{k+1}}=j|\mu_{t_k}=i)$, the joint likelihood of the data can be written as follow
\begin{eqnarray}
&\!\!\!& L(\alpha,\beta,\lambda_1,\lambda_2|m_{t_0},v_{t_0},m_{t_1},v_{t_1},\cdots,m_{t_K},v_{t_K})\nonumber\\
&\propto& f(m_{t_1},v_{t_1}|m_{t_0},v_{t_0},\alpha,\beta,\lambda_1,\lambda_2) \times \cdots  \times f(m_{t_K},v_{t_K}|m_{t_{K-1}},v_{t_{K-1}},\alpha,\beta,\lambda_1,\lambda_2),\nonumber
\end{eqnarray}
where $f(m_{t_0},v_{t_0})$ is assumed to be constant. For the assumptions that all the trajectories are independent and the number of cell population at initial state $N_0\gg1$, then $R^X(t)\sim N(\mu_t,\sigma_{t}^2)$ as $N_0 \rightarrow +\infty $ according to the asymptotic normality of $R^X(t)$ by \cite{yakovlev2009RelativeFrequenciesMultitype}, so the asymptotic distributions of $M_t$ and $V_t$ is given by:
\begin{eqnarray}
	&M_t\sim N(\mu_t,\frac{\sigma_{t}^2}{n}),
	\quad V_t\sim \frac{\sigma_{t}^2}{n-1}\chi^2(n-1),&\nonumber
\end{eqnarray}
here $M_t$ and $V_t$ are independent, $n$ is the number of trajectories. Then the joint distribution of the sample mean and sample variance at the next moment $f(m_{t_{k+1}},v_{t_{k+1}})$ can be written as the product of two independent distributions $f(m_{t_{k+1}})$ and $f(v_{t_{k+1}})$, i.e.,
	\begin{normalsize}
		\begin{eqnarray}
			&\enspace&f(m_{t_{k+1}},v_{t_{k+1}})\nonumber\\
			&=&f(m_{t_{k+1}})\times f(v_{t_{k+1}})\nonumber\\
&=&\!\!\frac{\sqrt{n}}{\sqrt{2\pi}\sigma_{t_{k\!+\!1}}}\!\!\times\! exp\!\left\{\frac{\!-n(m_{t_{k\!+\!1}}\!-\!\mu_{t_{k\!+\!1}})^2}{2\sigma^2_{t_{k\!+\!1}}}\right\} \times\left(\frac{n\!-\!1}{\sigma^2_{t_{k\!+\!1}}}\right)^{\frac{n\!-\!1}{2}}\!\!\!\!\!\!\!\times\!\! \frac{v_{t_{k\!+\!1}}^{\frac{n\!-\!1}{2}\!-\!1}}{2^{\frac{n\!-\!1}{2}}\Gamma(\frac{n\!-\!1}{2})} 
			\!\!\times\!\! exp\left\{\frac{-(n\!-\!1)v_{t_{k\!+\!1}}}{2\sigma^2_{t_{k\!+\!1}}} \right\}.\nonumber\\
		\end{eqnarray}
	\end{normalsize}
	For $\mu_{t_{k+1}}$ and $\sigma^2_{t_{k+1}}$, they can be replaced by Equation~(\ref{eq:different ode}), and the expectation and variance at the previous moment, i.e. $\mu_{t_{k}}$ and $\sigma^2_{t_{k}}$, can be estimated by sample mean and sample variance at previous moment, i.e. $m_{t_{k}}$ and $v_{t_{k}}$. Hence,
		\begin{align}\label{eq:conditional distribution}
			 &\quad f(m_{t_{k+1}},v_{t_{k+1}}|m_{t_k},v_{t_k},\alpha,\beta,\lambda_1,\lambda_2)\nonumber\\		
		   &=\!\frac{\sqrt{n}}{\sqrt{2\pi\Sigma_{t_k,t_{k+1}}(\alpha,\beta,\lambda_1,\lambda_2,m_{t_k},v_{t_k},N_{t_{k}},N_{t_{k+1}})}}
        \times exp\left\{\frac{\!-n\left[m_{t_{k\!+\!1}}\!-\!A_{t_k}(\alpha,\beta,\lambda_1,\lambda_2,m_{t_k})\right]^2}
			{2\Sigma_{t_k,t_{k+1}}(\alpha,\beta,\lambda_1,\lambda_2,m_{t_k},v_{t_k},N_{t_{k}},N_{t_{k+1}})}\right\} \nonumber \\
			&\times\!\!\left[\!\frac{n\!-\!1}{\Sigma_{t_k,t_{k+1}}\!(\alpha,\!\beta,\!\lambda_1,\!\lambda_2,\!m_{t_k},\!v_{t_k},\!N_{t_{k}},\!N_{t_{k+1}}\!)} \!\!\right]^{\!\!\frac{n\!-\!1}{2}}
			\!\!\!\!\!\!\!\!\!\times\!\frac{v_{t_{k\!+\!1}}^{\frac{n\!-\!3}{2}}}{2^{\frac{n\!-\!1}{2}}\Gamma(\frac{n\!-\!1}{2})}\times\!\! exp\!\left\{\!\frac{-(n\!-\!1)v_{t_{k\!+\!1}}}{2\Sigma_{t_k,t_{k+1}}\!(\alpha,\!\beta,\!\lambda_1,\!\lambda_2,\!m_{t_k},\!v_{t_k},\!N_{t_{k}},\!N_{t_{k+1}}\!)}\!\!\right\}\!\!.\nonumber\\
\end{align}
%
	Because even the most rapidly dividing human cells require at least one day to complete a proliferation cycle \citep{cowan2004DerivationEmbryonicStemCell}, if the number of cells is assumed to grow exponentially, it is easy to know that the fastest growth rate is $ln2$. So non-informative priors $\alpha,\beta\sim Unif(0,1),\lambda_1,\lambda_2\sim Unif(0,ln2)$ are applied. Set $\theta=\left\{\alpha,\beta,\lambda_1,\lambda_2\right\}$, the posterior distribution is obtained as follows
	\begin{eqnarray}
		&\quad&p(\theta|m_{t_0},v_{t_0},\cdots,m_{t_K},v_{t_K})\nonumber\\
		&=&\frac{p(m_{t_0},v_{t_0},\cdots,m_{t_K},v_{t_K}|\theta)\times p(\theta)}{\int_{\theta}
			p(m_{t_0},v_{t_0},\cdots,m_{t_K},v_{t_K}|\theta)\times p(\theta)d\theta} \nonumber\\
		&\propto&L(\theta|m_{t_0},v_{t_0},\cdots,m_{t_K},v_{t_K})p(\theta)\nonumber\\
		&\propto& \prod\limits_{k=0}^{K-1} f(m_{t_{k+1}},v_{t_{k+1}}|m_{t_k},v_{t_k},\theta)p(\theta)\nonumber\\
		&\propto& \prod\limits_{k=0}^{K-1} \Sigma_{t_k,t_{k+1}}^{-\frac{n}{2}}(\theta,m_{t_k},v_{t_k},N_{t_{k}},N_{t_{k+1}}) v_{t_{k\!+\!1}}^{\frac{n\!-\!3}{2}} \times exp \Bigg\{\!\!\!-\!\frac{1}{2} \sum\limits_{k=0}^{K-1}\!
		\bigg\{\frac{n\left[m_{t_{k\!+\!1}}\!\!-\!\!A_{t_k}(\theta,m_{t_k})\right]^2\!\!+\!(n\!-\!1)v_{t_{k\!+\!1}}}{\Sigma_{t_k,t_{k+1}}(\theta,m_{t_k},v_{t_k},N_{t_{k}},N_{t_{k+1}})}\!\bigg\}\!
		\!\Bigg\}.\nonumber
	\end{eqnarray}
	Here $N_{t_k}$ can be obtained by Equation~(\ref{eq:discrete Nt}).
	Actually, this framework can be applied in different data types. For example, if every trajectory of CSCs proportion was recorded, then $m_{t_k}$ in Equation~(\ref{eq:conditional distribution}) can be estimated by the proportion of CSCs at that time, and $v_{t_k}$ in Equation~(\ref{eq:conditional distribution}) can be treated as zero. Hence the posterior distribution of parameters can be obtained analogously (See Web Appendix B for more details).
\subsection{Algorithm}\label{sub:algorithm}
	It is well known that, the smaller $\Delta t$ in Equation~(\ref{eq:different ode}), the better accuracy of discretization. More importantly, if $\Delta t$ is too big, the variance in difference equation (\ref{eq:different ode}) may be negative. However, the real dataset applied is recorded every second day, i.e., $\Delta t =2$ days, which is too large to deduce $\sigma^2_{t+1}$ precisely. This problem can be resolved by imputing missing data. That is, the data can be considered as the following form
	\begin{align*}
		\textbf{m}_{obs}&=\left\{m_{t_0},m_{t_1},\cdots,m_{t_K}\right\}\\
		\textbf{v}_{obs}&=\left\{v_{t_0},v_{t_1},\cdots,v_{t_K}\right\} \\
		\textbf{m}_{mis}&=\left\{m_\frac{2t_0+t_1}{3},m_\frac{t_0+2t_1}{3},\cdots,m_{\frac{2t_{K-1}+t_K}{3}},m_{\frac{t_{K-1}+2t_K}{3}}\right\}\\
		&\triangleq \left\{m_{t^*_{1\slash3}},m_{t^*_{2\slash3}},\cdots,m_{t^*_{K\!-\!2\!\slash\!3}},m_{t^*_{K\!-\!1\!\slash\!3}}\right\}\\
		\textbf{v}_{mis}&=\left\{v_\frac{2t_0+t_1}{3},v_\frac{t_0+2t_1}{3},\cdots,v_{\frac{2t_{K-1}+t_K}{3}},v_{\frac{t_{K-1}+2t_K}{3}}\right\}\\
		&\triangleq \left\{v_{t^*_{1\slash3}},v_{t^*_{2\slash3}},\cdots,v_{t^*_{K\!-\!2\!\slash\!3}},v_{t^*_{K\!-\!1\!\slash\!3}}\right\}.
	\end{align*}
	Here define
\begin{eqnarray*}
		t^*_{k\!+\!1\!\slash\!3}=\frac{2t_k+t_{k+1}}{3},\enspace t^*_{k\!+\!2\!\slash\!3}=\frac{t_k+2t_{k+1}}{3},\enspace
		k=0,1,\cdots,K-1.
	\end{eqnarray*}
	The $\textbf{m}_{mis}$ and $\textbf{v}_{mis}$ represent the unobserved data, with assumption that two records are missing between every two observations, i.e. $\Delta t$ is modified as $\frac{2}{3}$. Then, the posterior distribution can be rewritten as:
\begin{small}
		\begin{eqnarray}
			&&\!\!\!p(\theta,\textbf{m}_{mis},\textbf{v}_{mis}|\textbf{m}_{obs},\textbf{v}_{obs})\nonumber\\
			&\propto&\!\!\!\!\!\prod\limits_{k=0}^{K-1}\!\!\left[\!(\Sigma_{t_k,t^*_{k+1\!\slash\!3}}\Sigma_{t^*_{k+1\!\slash\!3},t^*_{k+2\!\slash\!3}}
			\Sigma_{t^*_{k+2\!\slash\!3},t_{k+1}}
			)^{-\!\frac{n}{2}}
			(v_{t^*_{k+ 1\!\slash\!3}}v_{t^*_{k+ 2\!\slash\!3}}v_{t_{k\!+\!1}})^{\frac{n-3}{2}}\!\right]\nonumber\\
			&\times&\!\!\!\!\! exp\!\left\{\!\!-\frac{1}{2}\!\sum\limits_{k=0}^{k-1}\!\!\left[
			\frac{n(m_{t^*_{k+ 1\!\slash\!3}}\!\!\!-\!\!A_{t_k})^2\!\!+\!\!(n\!-\!1)v_{t^*_{k+ 1\!\slash\!3}}}{\Sigma_{t_k,t^*_{k+1\!\slash\!3}}}
			\!+\!\frac{n(m_{t^*_{k+ 2\!\slash\!3}}\!\!\!-\!\!\!A_{t^*_{k+ 1\!\slash\!3}})^2\!\!+\!\!(n\!\!-\!\!1)v_{t^*_{k+ 2\!\slash\!3}}}{\Sigma_{t^*_{k+1\!\slash\!3},t^*_{k+2\!\slash\!3}}}
			\!+\!\frac{n(m_{t_{k+1}}\!\!\!-\!\!\!A_{t^*_{k+2\!\slash\!3}})^2\!\!+\!\!(n\!\!-\!\!1)v_{t_{k\!+\! 1}}}{\Sigma_{t^*_{k+2\!\slash\!3},t_{k+1}}}\!\right]\!\!\right\}\!\!.\nonumber
		\end{eqnarray}
	\end{small}
Here $\theta$ is omitted for simplicity. The conditional posterior distributions of all parameters and missing data are shown in Web Appendix C.
As to $N_t$, due to $\mu_s$ is unknown, it can be estimated by splitting and summing
	\begin{eqnarray*}
		&N_{t_k}\approx N_0e^{\sum\limits_{q=0}^{k-1} \frac{m_{t_q}+m_{t_{q+1}}}{2}(3\Delta t)+\lambda_2 t_k}&\\
		&N_{t^*_{k+1\!\slash\!3}}\approx N_0e^{\sum\limits_{q=0}^{k-1} \frac{m_{t_q}+m_{t_{q+1}}}{2}(3\Delta t)+\frac{m_{t_k}+m_{t^*_{k+1\!\slash\!3}}}{2}\Delta t+\lambda_2 t_k}&\\
		&N_{t^*_{k+2\!\slash\!3}}\approx N_0e^{\sum\limits_{q=0}^{k-1} \frac{m_{t_q}+m_{t_{q+1}}}{2}(3\Delta t)+\frac{m_{t_k}+m_{t^*_{k+2\!\slash\!3}}}{2}(2\Delta t)+\lambda_2 t_k}.&
	\end{eqnarray*}

Then, standard MCMC methods with Gibbs Sampling \citep{geman1984StochasticRelaxationGibbs} and MH algorithm \citep{metropolis1953EquationStateCalculations} was applied to draw four MCMC chains independently and the first half of every chain are abandoned as pre-burned. Multivariate Potential Scale Reduction Factor (MPSRF) $\hat{R}$ \citep{gelman1992InferenceIterativeSimulation,brooks1998GeneralMethodsMonitoring} was used to judge the convergence of MCMC chains. When $\hat{R}< 1.1$, we consider MCMC chains as convergent, the mean value and interval between $2.5\% $ and $97.5\%$ quantiles of the converged posterior samples are acquired as the point estimation and interval estimation respectively.

\section{Results}\label{sec:Results}
 In this section, we perform several types of simulations and also apply our method to real data. Simulation I and II adopt different data generation methods to demonstrate the validity and efficacy of the algorithm, while Simulation II conducts a comparison among our proposed method, the NLS method and the discrete time model, as the other associated methods demand a substantial volume of data or intricate optimization algorithms. In Simulation III, the impact of different initial cell counts $N0$ on the estimation outcomes are investigated. Moving towards practical applications with real data, our method is employed to analyze data derived from the SW620 colon cancer cell line, as outlined in Section 2.1. This practical application offers insights into performance of algorithm in a real data context.

\subsection{Simulation I}\label{Simulation I}

In this part, in order to verify our algorithm, we generated synthesized datasets
\[
	\left\{(m_0,v_0),(m_\frac{2}{3},v_\frac{2}{3}),(m_\frac{4}{3},v_\frac{4}{3}),
	(m_2,v_2),(m_\frac{8}{3},v_\frac{8}{3}),(m_\frac{10}{3},v_\frac{10}{3}),(m_4,v_4),\cdots,(m_{24},v_{24})\right\}
\]
sequentially based on the conditional distribution Equation~(\ref{eq:conditional distribution}) in Section \ref{sub:Bayesian framework}. Let $t=24, \Delta t =\frac{2}{3}$, assuming no missing value. To be consistent with real data, we set the number of trajectory $n=5$ and initial data $N_0=1000$, which follows previous work \cite{wang2014DynamicsCancerCell}. We generated 100 groups of parameters from the prior, i.e., initial mean value $m_0\sim Unif(0,1)$ and variance $v_0\sim Unif(0,0.01)$, $\alpha,\beta\sim Unif(0,1),\enspace \lambda_1,\lambda_2\sim Unif(0,ln2)$.

Our method is applied to estimate the parameters. Table~\ref{tab:tabone} shows the estimation results including Average Square Error (ASE) of point estimation, and mean proportion of interval estimation coverage (CR). It shows that our algorithm is accurate for the synthesized data, with small ASE and high coverage rate.

\begin{table}
	\caption{Results of Simulation I and II\label{tab:tabone}}
	\begin{center}
		\begin{tabular}{@{}ccccccc@{}}
			\hline
			\multicolumn{3}{c}{} & \multicolumn{1}{c}{$\alpha$}
			&\multicolumn{1}{c}{$\beta$} & \multicolumn{1}{c}{$\lambda_1$}
			&\multicolumn{1}{c}{$\lambda_2$}  \\
			\cline{1-7}
			\multirow{2}{*}{Simulation I}&\multirow{2}{*}{Proposed model} &ASE&0.0182&0.0165&0.0079&0.0081 \\  
			&&CR&0.96&0.96&0.96&0.95\\ \hline
			\multirow{7}{*}{Simulation II}&\multirow{2}{*}{Proposed model} &ASE&0.0250&0.0300&0.0174 &0.0181\\
			&&CR&0.90&0.88&0.89&0.88 \\  \cline{3-7}
            &\multirow{2}{*}{Discrete model}&ASE&0.0621&0.0626&\mbox{-} &\mbox{-}\\
            &&CR&0.21&0.22&\mbox{-}&\mbox{-}\\ \cline{3-7}
            &NLS1 &ASE  &0.0999&0.1000&0.0329&0.0360 \\ \cline{3-7}
            & NLS2&ASE &0.0741&0.873&0.0484&0.0582\\
        \hline
		\end{tabular}
	\end{center}
\end{table}

\subsection{Simulation II}
To enhance alignment with real data, we employed a distinct data generation approach in Simulation II, differing from that used in Simulation I. Additionally, missing data was introduced in this section. Specifically, for this simulation, we set $t=24$ and $\Delta t =\frac{2}{3}$, resulting in the generation of only 13 data points. This implies the presence of missing data that requires estimation, with two observations missed between each consecutive two days i.e.
\begin{eqnarray*}
	&\textbf{Z}_{obs}=\left\{(m_0,v_0),(m_2,v_2),(m_4,v_4),\cdots,(m_{24},v_{24})\right\}&\\ &\textbf{Z}_{mis}=\left\{(m_\frac{2}{3},v_\frac{2}{3}),(m_\frac{4}{3},v_\frac{4}{3}),(m_\frac{8}{3},v_\frac{8}{3}),(m_\frac{10}{3},v_\frac{10}{3}),\cdots,(m_\frac{68}{3},v_\frac{68}{3}),(m_\frac{70}{3},v_\frac{70}{3})\right\}.&
\end{eqnarray*}
In addition, the data was generated by Gillespie's algorithm \citep{gillespie1977ExactStochasticSimulation}, which simulates the cellular growth and state transition. In fact, the data generated in this way are more in line with the data obtained from actual cell growth. Initial mean value was set as $m_0\sim Unif(0,1)$, and then synthesized trajectories of $R^X(t)$ from the same initial proportion are generated. The calculated sample mean $m_t$ and sample variance $v_t$ are the input of our algorithm. The values of the interested parameters are generated as the same in Section \ref{Simulation I}, i.e., $\alpha,\beta\sim Unif(0,1),\enspace \lambda_1,\lambda_2\sim Unif(0,ln2)$ and $N_0=1000$. In total we generated 100 groups of parameters and then applied our method to the parameter estimation. The results are contained in Table~\ref{tab:tabone}. Comparing with the complete data case in Simulation I, ASE is a little larger and CR decreases slightly, but due to deviations between model and reality, it still proves that our method is effective when the data is missing partially.

At the same time, a comparative analysis involving our proposed method, the NLS method introduced in \cite{xue2010SieveEstimationConstant}, and the discrete model outlined in \cite{zhou2018BayesianStatisticalAnalysis} was conducted in this section. The discrete model represents a distinct branching process model for the same problem of CSC plasticity. Employing a Bayesian approach, we compare the ASE and the CR of this discrete model with our proposed method. On the other hand, the NLS method serves as an alternative for parameter estimation in the proposed ODEs (\ref{eq:keyODEs}). We consider two distinct objective functions for the NLS method: NLS1, defined as $\sum{(\mu_t-\hat{\mu_t})^2+(\sigma^2_t-\hat{\sigma}^2_t)^2}$, and NLS2, defined as $\sum{(\mu_t-\hat{\mu_t})^2+(\sigma_t-\hat{\sigma_t})^2}$.

The outcomes of this comparative study are presented in Table~\ref{tab:tabone}. We evaluated the ASE of point estimates for parameters across the four models, where discrete model has only two parameters. Notably, our model demonstrates better alignment with real-world scenarios, evident from its smaller ASE and higher CR compared to the discrete model. Furthermore, our approach outperforms the NLS method in terms of ASE, underscoring its optimality for parameter estimation in nonlinear ODEs.

\subsection{Simulation III}
\begin{figure}[h]
  \centerline{\includegraphics[width=15cm]{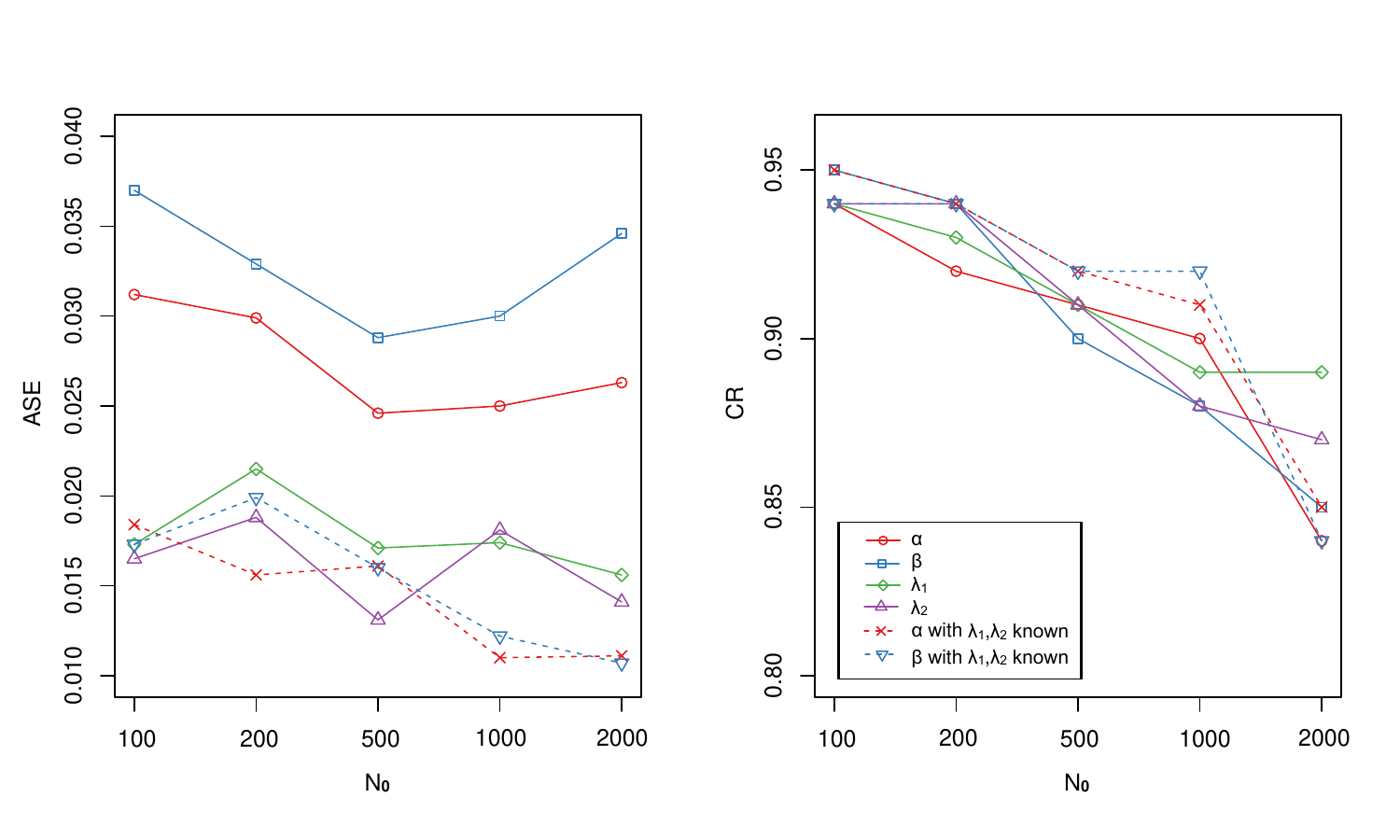}}
  \caption{The simulation results with different initial total number of cells $N_0$. The two panels show the ASE and CR of simulations respectively. The solid lines represent the estimated results for the four parameters at different initial values. The dash lines represent the results in the case where $\lambda_1$ and $\lambda_2$ are known in advance. Each simulation contains 100 groups of parameters.}
\label{fig:fifth}
\end{figure}

In order to assess the influence of $N_0$ on estimation accuracy, an extensive analysis considering various values of $N_0$ was conducted in this section. With Gillespie's algorithm, the data was generated independently with different $N_0$ values: $N_0=\{100,200,500,1000,2000\}$. The remaining parameters  were kept consistent with the aforementioned specifications. Furthermore, recognizing that the rate of cell division $\lambda$ can be directly measured in certain scenarios \citep{wang2010RobustGrowthEscherichia,tanouchi2015NoisyLinearMap}, we set $\lambda_1$ and $\lambda_2$ as true values, aiming to estimate the other two parameters and examine potential enhancements in performance. The findings are depicted in Figure~\ref{fig:fifth} and reveal intriguing insights.

The findings are depicted in Figure~\ref{fig:fifth} and reveal intriguing insights. Notably, CR experiences a significant decline as $N_0$ increases, while the ASE demonstrates relatively minor fluctuations. This pattern can be attributed to the increasing reliability of approximating $N_t$ as a deterministic variable as $N_0$ grows larger, resulting in a decrease in ASE at the initial stage. However, as $N_0$ further increases, the system attains equilibrium more swiftly, thereby reducing the amount of information captured within the generated data. Consequently, the estimation effectiveness diminishes, leading to the observed increase in ASE and decline in CR. These outcomes collectively suggest that our method can be optimally applied with experiments designed using a smaller initial cell count.

Furthermore, it is evident that if an independent measurement of the cell division rate is available a priori, our model showcases improved performance in estimating $\alpha$ and $\beta$, manifesting a significant reduction in ASE. This underscores the potential utility of such supplementary information for enhancing estimation accuracy.

\subsection{Real data analysis}

Our real data application is based on a published dataset derived from SW620 colon cancer cell line. With this dataset, two issues were considered, one is whether the hypothesis of reversible phenotypic plasticity  outperforms the hypothesis of cellular hierarchy, and the other is whether the two types of cells growth at the same rate. As a result, there are four competing models were considered as follow
\begin{center}
(1)$\beta=0,\lambda_1=\lambda_2$;
(2)$\beta\neq0,\lambda_1=\lambda_2$;
(3)$\beta=0,\lambda_1\neq\lambda_2$;
(4)$\beta\neq0,\lambda_1\neq\lambda_2$.
\end{center}
To evaluate these models, our method was applied to four distinct sets of real data, differing in the initial states of relative frequencies. The Deviance Information Criterion (DIC) \citep{gelman2015BayesianDataAnalysis} was employed to facilitate model selection.
\begin{table}[t]
\centering
\Huge
\setlength{\belowcaptionskip}{0.2cm}
\caption{Results of Real Data\label{real data1}}
\resizebox{1.05\columnwidth}{!}{
	\begin{tabular}{@{}ccccccccccccccccccccc@{}}
		\hline
		\multirow{2}{*}{}&& \multicolumn{3}{c}{$\beta=0,\lambda_1=\lambda_2$}&
		& \multicolumn{4}{c}{$\beta\neq0,\lambda_1=\lambda_2$}&
		& \multicolumn{4}{c}{$\beta=0,\lambda_1\neq\lambda_2$}& &\multicolumn{5}{c}{$\beta\neq0,\lambda_1 \neq \lambda_2$} \\ \cline{3-5}\cline{7-10} \cline{12-15}\cline{17-21}
		&&$\alpha$&$\lambda$&DIC&&$\alpha$&$\beta$&$\lambda$&DIC&&$\alpha$&$\lambda_1$&$\lambda_2$&DIC&&$\alpha$&$\beta$&$\lambda_1$&$\lambda_2$&DIC\\ \hline
		Group 1&&0.995&0.639&-195.050&&0.908 &0.200&0.613&-229.991& &0.972&0.647&0.545&-195.753&&0.947 &0.255&0.474&0.608 &\bf{-234.356}\\
		Group 2&&0.993&0.600&-287.651&&0.983 &0.043 &0.593&-297.119& &0.973&0.608&0.549&-281.338&&0.971 &0.047&0.586&0.564 &\bf{-301.963} \\
		Group 3&&0.941&0.561&-285.782&&0.787 &0.416 &0.424&-297.142 &&0.770&0.498&0.141&-327.872&&0.731 &0.309 &0.431&0.086&\bf{-334.087}\\
		Group 4&&0.851&0.111&-356.755&&0.736 &0.483 &0.056&-379.372 &&0.876&0.228&0.162&-377.511&&0.747 &0.453 &0.103&0.096 &\bf{-385.092}\\ \hline
\end{tabular}}
\end{table}

The outcomes of parameter estimation and model selection are presented in Table~\ref{real data1}, and the smallest DIC values are marked in bold. The parameter estimation results reveal the influence of varying initial states on parameter values. In terms of model selection, a consistent trend emerges: across scenarios where the division rates of the two cell types are either equal or different, the DIC values for the model $\beta\neq 0$ consistently outperform its counterpart. This consistent lower DIC suggests that incorporating de-differentiation enhances the model's fit to the provided data. This observation aligns seamlessly with the biological evidence presented in \cite{li2019ExosomalFMR1AS1Facilitates}.

Furthermore, it is notable that the model accounting for distinct cell division rates yields superior performance, implying heterogeneous growth rates among different cancer cell types. This result resonates with \cite{raatz2021ImpactPhenotypicHeterogeneity}, wherein growth rate-dependent treatments lead to delayed relapse compared to growth rate-independent treatments. Collectively, these findings highlight the variability in growth rates among diverse cancer cell populations.

\section{Discussions}\label{sec:discussions}

A Bayesian statistical analysis aimed at unraveling phenotypic plasticity within cancer cells utilizing temporal cell proportion data from cancer cell line is presented.  In contrast to conventional approaches that focus on compositional data, our methodology is based on a kinetic model that encapsulates the dynamic interactions between CSCs and NSCCs, with the moment equations governing the mean and variance of CSCs proportion derived. And a statistical analysis of temporal compositional data is developed to estimate the parameters in the moment equations.
Notably, the detection of phenotypic plasticity is reframed as the inference of whether the probability of de-differentiation from NSCCs to CSCs is bounded away from zero. When applied to the SW620 colon cancer cell line, our approach yields statistical inference results that align with \emph{in situ} experiments, affirming the capacity for NSCCs to de-differentiate into CSCs.

Indeed, our method holds potential for broader applications across diverse fields involving temporal compositional data. One notable example is the realm of microbiome research, where microbial data can be inherently treated as compositions \citep{knight2018BestPracticesAnalysing}. By leveraging our methodology, when microbiome data is collected over subsequent time points, our approach could yield more nuanced and detailed insights.
Beyond microbiomics, other fields such as environmental science \citep{sanchez-balseca2020SpatiotemporalAirPollution} and geochemical studies \citep{wang2021CompositionalDataAnalysis} also present scenarios characterized by compositional data. Given the increasing availability of temporal compositional data across various domains, the versatility of our method opens the door to a wide range of potential applications in the future.

In general, there are two promising directions for extending our approach. One is how to detect the dynamic switching patterns between CSCs and NSCCs. Given the ever-changing microenvironments within which these cells operate, the division patterns of cancer cells may exhibit temporal variability. \citep{ahmed2018microenvironment}. Rather than assuming stability, parameters like symmetric division probability and asymmetric division probability should be considered as time-varying. It is of great interest to develop robust statistical methods to identify and potentially quantify these time-variant patterns among different cell types. The other direction is to infer phenotypic plasticity based on \emph{in vivo} experiments. Our approach so far is devised to analyze \emph{in vitro} experiments (i.e. cancer cell line). More complex cell-to-cell interactions in tumor tissue such as nonlinear feedbeck control \citep{rodriguez2011evolutionary} are not taken into account. Furthermore, the consideration of more than two types of cancer cells becomes imperative in this context. To address these challenges, future research could focus on developing statistical methodologies that accommodate these complexities, utilizing \emph{in vivo} immunohistochemical experimental datasets to discern and characterize diverse cancer cell populations within tumors. Such endeavors would pave the way for a deeper understanding of complex cellular dynamics in real-world scenarios.

\newpage
\bibliographystyle{IEEEtran}
\bibliography{reference}

\begin{appendix}
\section*{Appendix A: Derivation of Equations(3)and(6)}\label{app:A}
The deduction of ODE on variance of the proportion of CSCs is as follow
\begin{eqnarray*}
	\frac{d\textbf{E}X^2_t}{dt}
	&=&\sum_{x,y} \frac{x^2 dP(x,y,t)}{dt} \nonumber\\
	&=&\lambda_1\alpha \sum_{x,y} P(x-1,y,t)x^2(x-1)+\lambda_2\beta\sum_{x,y} P(x-1,y,t)x^2 y \nonumber\\
	&\quad&+\lambda_1(1-\alpha)\sum_{x,y} P(x,y-1,t)x^3+\lambda(1-\beta) \sum_{x,y} P(x,y-1,t)x^2(y-1) \nonumber\\
	&\quad&-\sum_{x,y} P(x,y,t)(x\lambda_1+y\lambda_2)x^2 \nonumber\\
	&=&\lambda_1\alpha \sum_{x,y} P(x,y,t)x(x+1)^2+\lambda_2\beta\sum_{x,y} P(x,y,t)(x+1)^2 y \nonumber\\
	&\quad&+\lambda_1(1-\alpha)\sum_{x,y} P(x,y,t)x^3+\lambda_2(1-\beta) \sum_{x,y} P(x,y,t)x^2 y \nonumber\\
	&\quad&-\sum_{x,y} P(x,y,t)(x^3\lambda_1+x^2 y\lambda_2) \quad \mbox{(variable substitution)}\ \nonumber\\
	&=&\lambda_1\alpha (\textbf{E}X_t+2\textbf{E}X^2_t)+\lambda_2\beta (\textbf{E}Y_t+2\textbf{E}X_tY_t).
\end{eqnarray*}
Taking the derivative of Equation~(5) in the main paper, we can get
\begin{eqnarray*}
	\frac{d\textbf{E}X^2_t}{dt}
	&=&\frac{d(\sigma^2_t+\mu^2_t)N^2_t}{dt} \nonumber\\
	&=&\frac{d(\sigma^2_t N^2_t)}{dt}+\frac{d(\mu^2_t N^2_t)}{dt} \nonumber\\
	&=&N^2_t \frac{d\sigma^2_t}{dt}+2\sigma^2_t N_t\frac{dN_t}{dt}+2\mu^2_t N_t\frac{dN_t}{dt}+2N^2_t \mu_t\frac{d\mu_t}{dt}\nonumber
\end{eqnarray*}
\begin{eqnarray*}
	\Longrightarrow &\quad& \lambda_1\alpha (\textbf{E}X_t+2\textbf{E}X^2_t)+\lambda_2\beta (\textbf{E}Y_t+2\textbf{E}X_tY_t) \nonumber\\
	&=&N^2_t \frac{d\sigma^2_t}{dt}+2\sigma^2_t N_t\frac{dN_t}{dt}+2\mu^2_t N_t\frac{dN_t}{dt}+2N^2_t \mu_t\frac{d\mu_t}{dt}
\end{eqnarray*}
\begin{eqnarray*}\label{second moment}
	\Longrightarrow &\quad& 2\lambda_1\alpha \textbf{E}(\frac{X^2_t}{N^2_t})+\frac{1}{N_t}\lambda_1\alpha\mu_t+2\lambda_2\beta \textbf{E}(\frac{X_t}{N_t}\frac{Y_t}{N_t})+\frac{1}{N_t}\lambda_2\beta(1-\mu_t)\nonumber\\
	&=&\frac{d\sigma^2_t}{dt}+2\frac{1}{N_t}\frac{dN_t}{dt} \sigma^2_t +2\frac{1}{N_t}\frac{dN_t}{dt}\mu^2_t+2 \mu_t\frac{d\mu_t}{dt},
\end{eqnarray*}
substituting Equations~(4), (5) in the main paper and
\begin{eqnarray*}
	\textbf{E}(\frac{X_t}{N_t}\frac{Y_t}{N_t})&=&\textbf{E}(\frac{X_t}{N_t}(1-\frac{X_t}{N_t}))=\textbf{E}(\frac{X_t}{N_t}-\frac{X^2_t}{N^2_t})=\mu_t-(\sigma^2_t+\mu^2_t)
\end{eqnarray*} into Equations~(15) in the main paper, we get
\begin{eqnarray*}
	&\quad&2\lambda_1\alpha (\sigma^2_t+\mu^2_t)+\frac{1}{N_t}\lambda_1\alpha\mu_t+2\lambda_2\beta [\mu_t-(\sigma^2_t+\mu^2_t)]+\frac{1}{N_t}\lambda_2\beta(1-\mu_t)\nonumber\\
	&=&\frac{d\sigma^2_t}{dt}+2\frac{1}{N_t}\frac{dN_t}{dt} \sigma^2_t +2\frac{1}{N_t}\frac{dN_t}{dt}\mu^2_t+2 \mu_t[(\lambda_1\alpha-\lambda_2\beta-\frac{1}{N_t}\frac{dN_t}{dt})\mu_t+\lambda_2\beta].
\end{eqnarray*}
Finally, we have ODE on $\sigma_t^2$:
\begin{eqnarray*}
	\frac{d\sigma^2_t}{dt}=2(\lambda_1\alpha-\lambda_2\beta-\frac{1}{N_t}\frac{dN_t}{dt})\sigma^2_t+\frac{\lambda_1\alpha-\lambda_2\beta}{N_t}\mu_t+\frac{\lambda_2\beta}{N_t}.
\end{eqnarray*}

\section*{Appendix B: Bayesian Framework for Trajectory Data}\label{app:B}
Given the time-series data on relative frequencies of CSCs as follows
\[
\left(\begin{array}{cccc}%
r_{1t_0}&r_{1t_1}& \cdots& r_{1t_K}\\
	r_{2t_0}&r_{2t_1}& \cdots& r_{2t_K}\\
	\vdots &\vdots& \ddots &\vdots \\
	r_{nt_0}&r_{nt_1}& \cdots& r_{nt_K}
\end{array}
\right).
\]

where $r_{it_k}$ denotes the value of $R^X(t_k)$ in the $i^{th}$ trajectory. Similarly, all the trajectories are sampled independently and the number of cell population at initial state $N_0\gg1$. As for the same reason in Section 2.4 in the main paper, we have $R^X(t)\sim N(\mu_t,\sigma_{t}^2)$. By using the Markov property of the model, we can write the joint likelihood of the data as follows
\begin{eqnarray*}
	&\quad& L(\alpha,\beta,\lambda_1,\lambda_2|r_{1t_0},\cdots,r_{1t_K},\cdots,r_{nt_0},\cdots,r_{nt_K})\\
	&=&\prod\limits_{i=1}^{n} f(r_{it_0})\times f(r_{it_1}|r_{it_0},\alpha,\beta,\lambda_1,\lambda_2)\times \cdots\times f(r_{it_K}|r_{it_{K-1}},\alpha,\beta,\lambda_1,\lambda_2),
\end{eqnarray*}
where $f(r_{it_0})$ is assumed as constant, the likelihood can be written as
\begin{eqnarray*}
	&\quad& L(\alpha,\beta,\lambda_1,\lambda_2|r_{1t_0},\cdots,r_{1t_K},\cdots,r_{nt_0},\cdots,r_{nt_K}) \\
	&\propto& \prod\limits_{i=1}^{n} f(r_{it_1}|r_{it_0},\alpha,\beta,\lambda_1,\lambda_2)\times \cdots\times f(r_{it_K}|r_{it_{K-1}},\alpha,\beta,\lambda_1,\lambda_2),\nonumber
\end{eqnarray*}
here
\begin{small}
\begin{eqnarray*}
	&\quad& f(r_{it_{k+1}}|r_{it_k},\alpha,\beta,\lambda_1,\lambda_2)\\
	&=&\frac{1}{\sqrt{2\pi}\sigma_{i(t_{k+1})}}\times exp\left\{-\frac{(r_{it_{k+1}}-\mu_{it_{k+1}})^2}{2\sigma_{it_{k+1}}^2}\right\}\\
	&=&\!\!\!\!\!\!\frac{1}{\sqrt{2\pi\Sigma_{it_k,it_{k+1}}(\alpha,\!\beta,\!\lambda_1,\!\lambda_2,\!\mu_{it_k},\!\sigma^2_{it_k},\!N_{it_{k}},\!N_{it_{k+1}})}}\!\!\times \! exp\!\left\{\frac{\!-\left[r_{it_{k+1}}-A_{it_{k}}(\alpha,\beta,\lambda_1,\lambda_2,\mu_{it_{k}})\right]^2}{2\Sigma_{it_k,it_{k+1}}
(\alpha,\beta,\lambda_1,\lambda_2,\mu_{it_k},\sigma^2_{it_k},N_{it_{k}},N_{it_{k+1}})}\!\right\}\!.
\end{eqnarray*}
\end{small}
It is reasonable to substitute $\mu_{i{t_k}}$ by $r_{i{t_k}}$ and set $\sigma_{i{t_k}}=0$, we get
\begin{small}
	\begin{eqnarray*} &\quad&L(\alpha,\beta,\lambda_1,\lambda_2|r_{1t_0},\cdots,r_{1t_K},\cdots,r_{nt_0},\cdots,r_{nt_K})\\
		&\propto&\!\!\!\!\!\! \prod\limits_{i=1}^{n} \! \prod\limits_{k=0}^{K-1} \Sigma^{-\frac{1}{2}}_{it_k,it_{k+1}}\!(\alpha,\!\beta,\!\lambda_1,\!\lambda_2,\!r_{it_k},\!0,\!N_{it_{k}},\!N_{it_{k+1}})
		\!\!\times \!\! exp\left\{\!-\frac{\left[r_{it_{k+1}}\!-\!A_{it_{k}}(\alpha,\!\beta,\!\lambda_1,\!\lambda_2,\!r_{it_{k}})\right]^2}
		{2\Sigma_{it_k,it_{k+1}}(\alpha,\beta,\lambda_1,\lambda_2,r_{it_k},0,N_{it_{k}},N_{it_{k+1}})}\right\}.
	\end{eqnarray*}
\end{small}
Combined with priors as the same in Section 2.4 in the main paper, and let $\theta=\left\{\alpha,\beta,\lambda_1,\lambda_2\right\}$, the following posterior distribution is obtained:
\begin{eqnarray*}
	&\quad& p(\theta|r_{1t_1},\cdots,r_{1t_K},\cdots,r_{nt_1},\cdots,r_{nt_K})\\
	&=&\frac{L(\theta,r_{1t_1},\cdots,r_{1t_K},\cdots,r_{nt_1},\cdots,r_{nt_K})\times p(\theta) }{\int_{\theta}
		L(\theta,r_{1t_1},\cdots,r_{1t_K},\cdots,r_{nt_1},\cdots,r_{nt_K})\times p(\theta)d\theta} \\
	&\propto& L(\theta,r_{1t_1},\cdots,r_{1t_K},\cdots,r_{nt_1},\cdots,r_{nt_K})p(\theta)\\
	&\propto& \prod\limits_{i=1}^{n}\! \prod\limits_{k=0}^{K-1}\! \Sigma^{- \frac{1}{2}}_{it_k,it_{k+1}}\!(\theta,r_{it_k},0,N_{it_{k}},N_{it_{k+1}})
	\!\!\times\!\! exp\left\{-\frac{1}{2}\!\sum\limits_{i=1}^{n}\! \sum\limits_{k=0}^{K-1}\!\frac{\left[r_{i(t_{k+1})}\!-\!A_{it_{k}}(\theta,r_{it_{k}})\right]^2}
	{\Sigma_{it_k,it_{k+1}}(\theta,r_{it_k},0,N_{it_{k}},N_{it_{k+1}})}\!\right\}\!.
\end{eqnarray*}

With such posterior distribution, conditional posterior can be deduced and sample chains can be drawn by standard MCMC method. Then, parameters estimation and model selection results can be gotten as the algorithm in Section 2.5 in the main paper.

\section*{Appendix C: Conditional Posterior Distributions}\label{app:C}

	\begin{eqnarray*}
		&\bullet& \alpha|\beta,\lambda_1,\lambda_2,\textbf{m}_{mis},\textbf{v}_{mis},\textbf{m}_{obs},\textbf{v}_{obs}\\
		&\propto&\!\!\!\!\! \prod\limits_{k=0}^{K-1}\!\! \left[\Sigma_{t_k,t^*_{k+1\!\slash\!3}}(\alpha)\Sigma_{t^*_{k+1\!\slash\!3},t^*_{k+2\!\slash\!3}}(\alpha)
		\Sigma_{t^*_{k+2\!\slash\!3},t_{k+1}}(\alpha)
		\right]^{-\!\frac{n}{2}}\!\!\!\!\times\! exp\!\left\{\!\!-\frac{1}{2}\sum\limits_{k=0}^{k-1}\!\!
		\left[\!
		\frac{n\!\left[m_{t^*_{k+ 1\!\slash\!3}}\!\!\!-\!\!A_{t_k}(\alpha)\right]^2\!\!\!+\!(n\!-\!1)v_{t^*_{k+ 1\!\slash\!3}}}{\Sigma_{t_k,t^*_{k+1\!\slash\!3}}(\alpha)}
		\right.\right.\nonumber\\
		&&\!\! \left.\left.
		\!+\frac{n\left[m_{t^*_{k+ 2\!\slash\!3}}\!\!\!-\!\!A_{t^*_{k+ 1\!\slash\!3}}(\alpha)\right]^2\!+\!(n\!-\!1)v_{t^*_{k+ 2\!\slash\!3}}}{\Sigma_{t^*_{k+1\!\slash\!3},t^*_{k+2\!\slash\!3}}(\alpha)}
		+\frac{n\left[m_{t_{k+1}}\!\!\!-\!\!A_{t^*_{k+2\!\slash\!3}}(\alpha)\right]^2\!+\!(n\!-\!1)v_{t_{k\!+\! 1}}}{\Sigma_{t^*_{k+2\!\slash\!3},t_{k+1}}(\alpha)}\right]\right\}\nonumber
	\end{eqnarray*}
	\begin{eqnarray*}
		&\bullet&\beta|\alpha,\lambda_1,\lambda_2,\textbf{m}_{mis},\textbf{v}_{mis},\textbf{m}_{obs},\textbf{v}_{obs}\\
		&\propto&\!\!\!\!\! \prod\limits_{k=0}^{K-1}\!\! \left[\Sigma_{t_k,t^*_{k+1\!\slash\!3}}(\beta)\Sigma_{t^*_{k+1\!\slash\!3},t^*_{k+2\!\slash\!3}}(\beta)
		\Sigma_{t^*_{k+2\!\slash\!3},t_{k+1}}(\beta)
		\right]^{-\!\frac{n}{2}}\!\!\!\!\times\! exp\!\left\{\!\!-\frac{1}{2}\sum\limits_{k=0}^{k-1}\!\!
		\left[\!
		\frac{n\!\left[m_{t^*_{k+ 1\!\slash\!3}}\!\!\!-\!\!A_{t_k}(\beta)\right]^2\!\!+\!(n\!-\!1)v_{t^*_{k+ 1\!\slash\!3}}}{\Sigma_{t_k,t^*_{k+1\!\slash\!3}}(\beta)}
		\right.\right.\nonumber\\
		&&\!\! \left.\left.
		\!+\frac{n\left[m_{t^*_{k+ 2\!\slash\!3}}\!\!\!-\!\!A_{t^*_{k+ 1\!\slash\!3}}(\beta)\right]^2\!+\!(n\!-\!1)v_{t^*_{k+ 2\!\slash\!3}}}{\Sigma_{t^*_{k+1\!\slash\!3},t^*_{k+2\!\slash\!3}}(\beta)}
		+\frac{n\left[m_{t_{k+1}}\!\!\!-\!\!A_{t^*_{k+2\!\slash\!3}}(\beta)\right]^2\!+\!(n\!-\!1)v_{t_{k\!+\! 1}}}{\Sigma_{t^*_{k+2\!\slash\!3},t_{k+1}}(\beta)}\right] \right\}\nonumber
	\end{eqnarray*}
	\begin{eqnarray*}
		&\bullet&\lambda_1|\alpha,\beta,\lambda_2,\textbf{m}_{mis},\textbf{v}_{mis},\textbf{m}_{obs},\textbf{v}_{obs}\\
		&\propto&\!\!\!\!\! \prod\limits_{k=0}^{K-1}\!\! \left[\!\Sigma_{t_k,t^*_{k+1\!\slash\!3}}(\lambda_1)\Sigma_{t^*_{k+1\!\slash\!3},t^*_{k+2\!\slash\!3}}(\lambda_1)
		\Sigma_{t^*_{k+2\!\slash\!3},t_{k+1}}(\lambda_1)\!
		\right]^{-\!\frac{n}{2}}\!\!\!\!\times\! exp\!\left\{\!\!-\frac{1}{2}\!\!\sum\limits_{k=0}^{k-1}\!\!
		\left[\!
		\frac{n\!\left[m_{t^*_{k+ 1\!\slash\!3}}\!\!\!\!-\!\!A_{t_k}(\lambda_1)\right]^2\!\!\!\!+\!\!(n\!-\!\!1)v_{t^*_{k+ 1\!\slash\!3}}}{\Sigma_{t_k,t^*_{k+1\!\slash\!3}}(\lambda_1)}
		\right.\right.\nonumber\\
		&&\!\! \left.\left.
		\!+ \frac{n\left[m_{t^*_{k+ 2\!\slash\!3}}\!\!\!-\!\!A_{t^*_{k+ 1\!\slash\!3}}(\lambda_1)\right]^2\!+\!(n\!-\!1)v_{t^*_{k+ 2\!\slash\!3}}}{\Sigma_{t^*_{k+1\!\slash\!3},t^*_{k+2\!\slash\!3}}(\lambda_1)}
		+\frac{n\left[m_{t_{k+1}}\!\!\!-\!\!A_{t^*_{k+2\!\slash\!3}}(\lambda_1)\right]^2\!+\!(n\!-\!1)v_{t_{k\!+\! 1}}}{\Sigma_{t^*_{k+2\!\slash\!3},t_{k+1}}(\lambda_1)}\right] \right\}\nonumber
	\end{eqnarray*}
	\begin{eqnarray*}
		&\bullet&\lambda_2|\alpha,\beta,\lambda_1,\textbf{m}_{mis},\textbf{v}_{mis},\textbf{m}_{obs},\textbf{v}_{obs}\\
		&\propto&\!\!\!\!\!\! \prod\limits_{k=0}^{K-1} \!\! \left[\Sigma_{t_k,t^*_{k+1\!\slash\!3}}(\lambda_2)\Sigma_{t^*_{k+1\!\slash\!3},t^*_{k+2\!\slash\!3}}(\lambda_2)
		\Sigma_{t^*_{k+2\!\slash\!3},t_{k+1}}(\lambda_2)
		\right]^{-\!\frac{n}{2}}\!\!\!\!\times\! exp\!\left\{\!\!\!-\frac{1}{2}\!\!\sum\limits_{k=0}^{k-1}\!\!
		\left[\!
		\frac{n\!\left[\!m_{t^*_{k+ 1\!\slash\!3}}\!\!\!\!-\!\!A_{t_k}(\lambda_2)\!\right]^2\!\!+\!\!(n\!-\!\!1)v_{t^*_{k+ 1\!\slash\!3}}}{\Sigma_{t_k,t^*_{k+1\!\slash\!3}}(\lambda_2)}
		\right.\right.\nonumber\\
		&&\!\! \left.\left.
		\!+\frac{n\left[m_{t^*_{k+ 2\!\slash\!3}}\!\!\!-\!\!A_{t^*_{k+ 1\!\slash\!3}}(\lambda_2)\right]^2\!+\!(n\!-\!1)v_{t^*_{k+ 2\!\slash\!3}}}{\Sigma_{t^*_{k+1\!\slash\!3},t^*_{k+2\!\slash\!3}}(\lambda_2)}
		+\frac{n\left[m_{t_{k+1}}\!\!\!-\!\!A_{t^*_{k+2\!\slash\!3}}(\lambda_2)\right]^2\!+\!(n\!-\!1)v_{t_{k\!+\! 1}}}{\Sigma_{t^*_{k+2\!\slash\!3},t_{k+1}}(\lambda_2)}\right]\right\}\nonumber
	\end{eqnarray*}
	\begin{eqnarray*}
		&\bullet& m_{t^*_{k+1\!\slash 3}}|\alpha,\beta,\lambda_1,\lambda_2,\textbf{m}^{-(k+1\!\slash\!3)}_{mis},
		\textbf{v}_{mis},\textbf{m}_{obs},\textbf{v}_{obs} \\
		&\propto&\!\!  \left[\Sigma_{t_k,t^*_{k+1\!\slash\!3}}(m_{t^*_{k+1\!\slash 3}})\Sigma_{t^*_{k+1\!\slash\!3},t^*_{k+2\!\slash\!3}}(m_{t^*_{k+1\!\slash 3}})\right]^{-\!\frac{n}{2}}\nonumber\\
		&&\!\times exp\!\left\{\!\!-\frac{1}{2}
		\left[
		\frac{n(m_{t^*_{k+ 1\!\slash\!3}}\!\!\!-\!\!A_{t_k})^2\!+\!(n\!-\!1)v_{t^*_{k+ 1\!\slash\!3}}}{\Sigma_{t_k,t^*_{k+1\!\slash\!3}}(m_{t^*_{k+1\!\slash 3}})}\!+\!\frac{n\left[m_{t^*_{k+2\!\slash\!3}}\!\!\!-\!\!A_{t^*_{k+ 1\!\slash\!3}}(m_{t^*_{k+1\!\slash 3}})\right]^2\!+\!(n\!-\!1)v_{t^*_{k+ 2\!\slash\!3}}}{\Sigma_{t^*_{k+1\!\slash\!3},t^*_{k+2\!\slash\!3}}(m_{t^*_{k+1\!\slash 3}})} \right] \right\}\nonumber
	\end{eqnarray*}
	\begin{eqnarray*}
		&\bullet& m_{t^*_{k+2\!\slash\!3}}|\alpha,\beta,\lambda_1,\lambda_2,\textbf{m}^{-(k+2\!\slash\!3)}_{mis},
		\textbf{v}_{mis},\textbf{m}_{obs},\textbf{v}_{obs} \\
		&\propto&\!\! \left[\Sigma_{t^*_{k+1\!\slash\!3},t^*_{k+2\!\slash\!3}}(m_{t^*_{k+2\!\slash\!3}})
		\Sigma_{t^*_{k+2\!\slash\!3},t_{k+1}}(m_{t^*_{k+2\!\slash\!3}})
		\right]^{-\!\frac{n}{2}}\nonumber\\
		&&\!\times exp\!\left\{\!\!-\frac{1}{2}
		\left[
		\frac{n(m_{t^*_{k+ 2\!\slash\!3}}\!\!\!-\!\!A_{t^*_{k+ 1\!\slash\!3}})^2\!+\!(n\!-\!1)v_{t^*_{k+ 2\!\slash\!3}}}{\Sigma_{t^*_{k+1\!\slash\!3},t^*_{k+2\!\slash\!3}}(m_{t^*_{k+2\!\slash\!3}})}
		+\frac{n\left[m_{t_{k+1}}\!\!\!-\!\!A_{t^*_{k+2\!\slash\!3}}(m_{t^*_{k+2\!\slash\!3}})\right]^2\!+\!(n\!-\!1)v_{t_{k\!+\! 1}}}{\Sigma_{t^*_{k+2\!\slash\!3},t_{k+1}}(m_{t^*_{k+2\!\slash\!3}})}\right] \right\}\nonumber
	\end{eqnarray*}
	\begin{eqnarray*}
		&\bullet & v_{t^*_{k+1\!\slash\!3}}|\alpha,\beta,\lambda_1,\lambda_2,\textbf{m}_{mis},
		\textbf{v}^{-(k+1\!\slash\!3)}_{mis},\textbf{m}_{obs},\textbf{v}_{obs}\\
		&\propto&\! \Sigma^{-\!\frac{n}{2}}_{t^*_{k+1\!\slash\!3},t^*_{k+2\!\slash\!3}}\!(v_{t^*_{k+1\!\slash\!3}})
		v^{\frac{n-3}{2}}_{t^*_{k+1\!\slash\!3}}
		\!\!\times\!\! exp\!\left\{\!\!-\frac{1}{2}\!\!
		\left[
		\frac{(n\!-\!1)v_{t^*_{k+ 1\!\slash\!3}}}{\Sigma_{t_k,t^*_{k+1\!\slash\!3}}}
		\!+\!\frac{n(m_{t^*_{k+ 2\!\slash\!3}}\!\!\!-\!\!A_{t^*_{k+ 1\!\slash\!3}})^2\!\!+\!\!(n\!-\!1)v_{t^*_{k+ 2\!\slash\!3}}}{\Sigma_{t^*_{k+1\!\slash\!3},t^*_{k+2\!\slash\!3}}(v_{t^*_{k+1\!\slash\!3}})} \right]\!\!\right\}\nonumber
	\end{eqnarray*}
	\begin{eqnarray*}
		&\bullet& v_{t^*_{k+2\!\slash\!3}}|\alpha,\beta,\lambda_1,\lambda_2,\textbf{m}_{mis},
		\textbf{v}^{-(k+2\!\slash\!3)}_{mis},\textbf{m}_{obs},\textbf{v}_{obs}\\
		&\propto&\! \Sigma^{-\!\frac{n}{2}}_{t^*_{k+2\!\slash\!3},t_{k+1}}\!(v_{t^*_{k+2\!\slash\!3}})
		v^{\frac{n-3}{2}}_{t^*_{k+2\!\slash\!3}}
		\!\!\times\!\! exp\!\left\{\!\!-\frac{1}{2}\!\!
		\left[
		\frac{(n\!-\!1)v_{t^*_{k+ 2\!\slash\!3}}}{\Sigma_{t^*_{k+1\!\slash\!3},t^*_{k+2\!\slash\!3}}}
		\!+\!\frac{n(m_{t_{k+1}}\!\!\!-\!\!A_{t^*_{k+ 2\!\slash\!3}})^2\!\!+\!\!(n\!-\!1)v_{t_{k+1}}}{\Sigma_{t^*_{k+2\!\slash\!3},t_{k+1}}(v_{t^*_{k+2\!\slash\!3}})} \right]\!\!\right\}\!.\nonumber
	\end{eqnarray*}
Here $\textbf{m}^{(-q)}_{mis}$ and $\textbf{v}^{(-q)}_{mis}$  denote $\textbf{m}_{mis}$ and $\textbf{v}_{mis}$ with $m_{t^*_q}$ and $v_{t^*_q}$ removed respectively, and parameters to be updated will be marked out in $A_{t_k}$ and $\Sigma_{t_k,t_{k+1}}$.

\end{appendix}

\end{document}